%% file: Asym_Tardos.tex
\documentclass{article}
\usepackage{spconf,amsmath,amssymb,graphicx,color,url}

\def\Buyer{B}
\def\Provider{P}
\def\listed{D_{j,i,k}=\textbf{CE}(S_j,(\pi_{j}(k)\|K_{\pi_{j}(k),i}))}
\def\listedrow{\mathcal{D}_{j,i}}
\def\U{U_{k(j,i),i}=\textbf{CE}(R_{j,i},D_{j,i,k(j,i)})}
\def\Urow{U_{k(j,i),i}}
\def\R{R_{j,i}}
\def\decrypt{\pi_{j}(k)\|K_{\pi_{j}(k),i} = \textbf{CE}^{-1}(\R,V)}
\def\listc{C_{\pi_{j}(k),i} = \textbf{E}(K_{\pi_{j}(k),i},O_{\pi_{j}(k),i})}
\def\listcrow{C_{\pi_{j}(k),i}}
\def\decryptC{O_{\pi_{j}(k),i}=\textbf{E}^{-1}(K_{\pi_{j}(k),i},C_{\pi_{j}(k),i})}

\def\longu{m}

\def\ind{\mathsf{ind}}

\definecolor{MyDarkGreen}{rgb}{0.1,0.55,0}

\graphicspath{{figures/}}

\title{An Asymmetric Fingerprinting Scheme based on Tardos Codes}
\name{Ana Charpentier$^{(a)}$, Caroline Fontaine$^{(b)}$, Teddy Furon$^{(a)}$, Ingemar Cox$^{(c)}$}
\address{$^{(a)}$INRIA-Rennes research center, Campus de Beaulieu, Rennes, France\\
$^{(b)}$CNRS/Lab-STICC/CID, T\'{e}l\'{e}com Bretagne/ITI, Brest, France\\
$^{(c)}$University College London, Dpt. of Computer Science, London, United Kingdom}

\begin{document}
%
\maketitle
\begin{abstract}

Tardos codes are currently the state-of-the-art in the design of
practical collusion-resistant fingerprinting codes.
Tardos codes rely on a secret vector drawn from a publicly known probability
distribution in order to generate each
Buyer's fingerprint.  For security purposes, this secret vector must not be revealed to
the Buyers.  To prevent an untrustworthy Provider forging a copy of a
Work with an innocent Buyer's fingerprint, previous asymmetric
fingerprinting algorithms enforce the idea of the Buyers generating their own
fingerprint. Applying this concept to Tardos codes is challenging since the
fingerprint must be based on this vector secret.

This paper provides the first solution for an
asymmetric fingerprinting protocol dedicated to Tardos codes.
The motivations come from a new attack, in which an
untrustworthy Provider by modifying his secret vector frames an innocent Buyer.

\end{abstract}
\begin{keywords}
Asymmetric fingerprinting, Tardos code
\end{keywords}
\section{Introduction}
This paper considers a problem arising in the fingerprinting of
digital content.  In this context, a fingerprint is a binary code that
is inserted into the Work for the purpose of protecting it from
unauthorized use, or, more precisely, for the purpose of identifying
individuals responsible for unauthorized use of a Work.
In such a scenario, it is assumed that two or more users
may collude in order to try to hide their identities.  In this case,
it is further assumed that colluders cannot alter those bits of the
code that are identical for all colluders.  However, where bits differ
across colluders, these bits may be assigned arbitrary values.  A key
problem is resistance to collusion, i.e. if a coalition of $c$ users
creates a pirated copy of the Work, its tampered fingerprint (i) should not
implicate innocent users, and (ii) should identify at least one of
the colluders.

This problem has received considerable attention since Boneh and
Shaw~\cite{Boneh1998Collusion} discussed the problem.
They first introduced the concept of totally $c$-secure codes:
if a coalition of $c$ users colludes to produce a
pirate copy of the Work, the tampered fingerprint is still guaranteed to identify at
least one of the colluders, with no chance of framing an innocent.
Boneh and Shaw showed that totally
$c$-secure binary codes do not exist for $c>1$. They then introduced the
concept of a $c$-secure code such that the probability of framing an innocent is lower than $\epsilon$.
Unfortunately, the length of
their codes, $O(c^4
\log(\frac{n}{\epsilon})\log(\frac{1}{\epsilon}))$, where $n$ is the
number of users, was such as to make them impractical.
Following Boneh and Shaw's paper, there has been considerable
effort to design shorter codes.

In 2003, Tardos~\cite{Tardos2003Optimal} proposed an efficient code
construction that, for the first time, reduced the code length
  to the lower bound, $O(c^2 \log(\frac{n}{\epsilon}))$, thereby
making such codes practical.  Tardos codes are currently the
state-of-the-art for collusion-resistant fingerprinting.

Several papers have considered a scenario where the Provider is untrustworthy.
Thanks to the knowledge of a Buyer's fingerprint,
the Provider creates a pirated copy of a Work, implicating this innocent Buyer.
To prevent this, Pfitzman \cite{PfitzmannAsym} first introduced the concept of asymmetric
fingerprinting in which the Provider doesn't need to know the Buyer's fingerprint. 
The Buyer first
commits to a secret (the fingerprint) that only he/she knows.  The
Buyer and Provider then follow a protocol which results in the Buyer
receiving a copy of the Work with his/her secret fingerprint (and some
additional information coming from the Provider) embedded within it.
The Provider did not learn the Buyer's secret, and cannot
therefore create a forgery.  Unfortunately, in the case of Tardos
codes, fingerprints must be drawn from a particular
probability distribution depending on a secret vector only known to the Provider.
Thus, previous asymmetric fingerprinting methods cannot be applied 
to Tardos codes.

The Tardos decoding is also vulnerable to an additional attack, in
which the Provider does {\em not} need to create a forgery.  Rather,
given any unauthorized copy, i.e. a Work that does {\em not} contain the
innocent Buyer's fingerprint, the Provider can alter its secret
vector in order to accuse an arbitrary Buyer.

Our paper is organized as follows. We briefly
  introduce Tardos codes in Sec.~\ref{sec:tardos}.  
  Sec.~\ref{sec:untrustworthy} describes the attack
  at the decoding side.  In order to
  prevent both the Buyer and the Provider from cheating, 
  Sec.~\ref{sec:Construction} presents a new asymmetric protocol
  specific to Tardos codes. Sec.~\ref{sec:wmaccuse} then discusses
  practical aspects of the fingerprints embedding and accusation.  We
  finally discuss our solution in Sec.~\ref{sec:discussions} before
  concluding.

\section{The Tardos fingerprinting code}
\label{sec:tardos}

For readers unfamiliar with Tardos codes, we now provide a brief
introduction. Further details can be found in~\cite{SkoricSymmetric}. 

Let $n$ denote the number of buyers, and $\longu$ the length of the
code. The fingerprints can then be arranged as a binary $n \times
\longu$ matrix $\mathbf{X}$, Buyer $j$ being related to the binary
fingeprint $\mathbf{X}_j = (X_{j1}, X_{j2}, \ldots , X_{j\longu})$.

To generate this matrix, $m$ real numbers $p_i\in[t,1-t]$ are
generated, each of them being randomly and independently drawn
according to the probability density function
$f:[t,1-t]\rightarrow\mathbb{R}^+$ with $f(z)=\kappa(t)
(z(1-z))^{-1/2}$ and $\kappa(t)^{-1}=\int_t^{1-t}(z(1-z))^{-1/2}dz$.
The parameter $t\ll 1$ is referred to as the cutoff. We set
$\mathbf{p}=(p_1,\ldots,p_{\longu})$. This vector $\mathbf{p}$ is the
secret key of the code only known by the Provider.
Each element of the matrix $\mathbf{X}$ is then
independently randomly drawn, such that the probability that an
element, $X_{ji}$, in the matrix is a one is given by
$\mathbb{P}(X_{ji} = 1) = p_i$. The fingerprint is then embedded
into the copy of the Work of the corresponding Buyer thanks to a watermarking technique.

If an unauthorized copy is found, its corresponding fingerprint,
$\mathbf{Y}$, is decoded.  Due to collusion, and possible distortions
such as transcoding, the decoded fingerprint is unlikely to exactly
equal one of the fingerprints in the matrix, $\mathbf{X}$.  To
determine if Buyer $j$ is involved in the production of the
unauthorized copy, a score, referred to as an accusation score, $S_j$
is computed.  If this score is greater than a given threshold $Z$,
then Buyer $j$ is considered to have colluded.

The scores are computed according to an accusation function $g$,
reflecting the impact of the correlation between the sequence
$\mathbf{X}_{j}$, associated with Buyer $j$, and the decoded sequence
$\mathbf{Y}$:
\begin{equation}
S_j = G(\mathbf{Y},\mathbf{X}_j,\mathbf{p}) = \sum^m_{i=1} g(Y_i,X_{ji},p_i).
\label{eq:AccusSum}
\end{equation}
In the usual symmetric codes~\cite{SkoricSymmetric}, function $g$ is
constrained (such that, for example, for an innocent, the expectation
of the score is zero and its variance is $m$), giving $g(1,1,p) =
g(0,0,1-p) = - g(0,1,p) = - g(1,0,1-p) = \sqrt{\frac{1-p}{p}}$.

\section{Untrustworthy content provider}
\label{sec:untrustworthy}

We now consider the case
where the Provider is no longer trusted, and, as such, wishes
to frame Buyer $j$.  In such a scenario, we assume that the Provider
has no prior access to an unauthorized copy, i.e. the Provider cannot
insert a false fingerprint into the unauthorized copy, nor can he/she
place a Buyer's copy on an unauthorized location.  On receipt of an
unauthorized copy, we further assume that the untrustworthy Provider
to extracts the corresponding fingerprint present in the unauthorized
copy.  We base this assumption on the hypothesis that the underlying
watermarking algorithm comes from a technology provider and that the
Provider doesn't master or has no access to this technology
brick.  Given the extracted fingerprint  $\mathbf{Y}$, the Provider must now compare
it to all known Buyers' fingerprints.  This comparison is performed
using Eq.~\eqref{eq:AccusSum}.  And it is here that the Provider can
lie, since the probabilities, $\mathbf{p}$, are only known to the
Provider.

An untrustworthy Provider can create a fake vector of probabilities,
$\hat{\mathbf{p}}$, that implicates Buyer $j$.  However, the
distribution, $f(p)$, is publicly known, so the question becomes, can
the Provider generate a $\hat{\mathbf{p}}$ that (i) implicates Buyer
$j$, and (ii) has an arbitrarily high probability of been drawn from the
distribution $f(p)$?

It is indeed extremely simple to do so. Let us focus on a column where
$p_{i}=p$ and $Y_{i}=X_{j,i}$. The true summand in
Eq.~\eqref{eq:AccusSum} is $g(1,1,p)$ or $g(0,0,p)$ (with equal
probability).  Suppose that the content provider replaces the secret
value $p$ by a fake secret $\hat{p}$ which is drawn independently
according to $f$. On average, this summand takes the new value:
$$
\Delta(t) = \int_{t}^{1-t}f(\hat{p})\frac{g(1,1,\hat{p})+g(0,0,\hat{p})}{2}d\hat{p}=\frac{1}{\pi}\ln\frac{1-t}{t}.
$$
For a cutoff $t=1/900$ (recommended by G. Tardos to fight against 3
colluders), the numerical value is surprisingly high:
$\Delta(1/900)\approx 2.16$.  Suppose now that the content provider
applies the same strategy on an index $i$ where $Y_{i}\neq
X_{j,i}$. Then the expectation is the opposite. However, in a Tardos code, even for an innocent Buyer $j$, the proportion
$\alpha$ of indices where symbols $Y_{i}$ and $X_{j,i}$ agree is above
$1/2$ for most of the collusion strategy.
For instance, with an interleaving collusion attack, $\alpha=3/4$ whatever the collusion size $c$.

Based on this knowledge, we propose the following attack. The Provider computes the score for
all Buyers, which, on average, equals 0 for innocent Buyers and
$2m/c\pi$ for the colluders~\cite{SkoricSymmetric}. The provider initializes
$\hat{\mathbf{p}}=\mathbf{p}$. Then, he/she randomly selects a column $i$
and randomly draws a fake secret $\hat{p_{i}}\sim f$. He/She re-computes
the score of Buyer $j$ with this fake secret and iterates selecting a
different column until $S_{j}$ is above the threshold $Z$.
On average, $m(c\pi\Delta(\alpha-1/2))^{-1}$ secret values $p_i$ need
to be changed in this way, e.g. only $20\%$ of the code length if the
copy has been made using an interleaving attack.

\begin{figure}
  \centering
  \includegraphics[width=\linewidth]{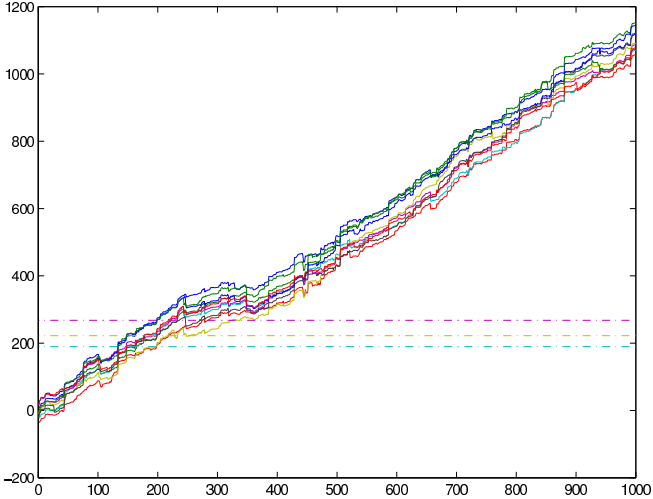}
  \caption{
Accusation score as a function of the number of changed elements of
the vector $\mathbf{p}$ for the case where $m=1000$ and $c=3$.
The solid coloured lines show how the accusation score of 10 randomly selected innocent buyers
increases as
more of the elements are modified.
The dotted horizontal lines show the original scores for the colluders
before the modification.}

  \label{fig:untrustworthy}
\end{figure}

Fig.~\ref{fig:untrustworthy} illustrates this attack for the case
where the code length is $m=1000$ and the
number of colluders is $c=3$.  The
solid coloured lines depict the accusation scores of 10 randomly selected innocent
buyers.  We observe that after between 20-30\% of the elements of $\mathbf{p}$
have been altered, the accusation scores of the innocent
Buyers exceed the {\em original} scores of the colluders.  In fact,
the colluders accusation scores also increase.  However, we are not
concerned with the highest score, but rather with any score
exceeding the threshold.  Thus, it is sufficient to raise the score of the
innocent Buyer, even if this raises all other Buyers' scores as well.

Randomly selecting some $p_{i}$'s (independently from $\mathbf{X}_j$
and $\mathbf{Y}$) and re-drawing them according to the same law
ensures that $\hat{p}_{i}\sim f$, $\forall i$.  Therefore, a judge
observing $\hat{\mathbf{p}}$ cannot distinguish the forgery.  For this
reason, the judge might request to see the matrix $\mathbf{X}$ to statistically test
whether the elements of $\mathbf{X}$ are drawn from the distribution
$\hat{\mathbf{p}}$. In this case, the
Provider can give a fake matrix $\hat{\mathbf{X}}$ where the columns
whose $p_{i}$ have been modified are re-drawn such that
$\mathbb{P}(X_{ki} = 1) = \hat{p}_i$, $\forall k\neq j$. 
The only way to prevent this deception would be if the
judge asked an innocent User $k\neq j$ for his copy in order to verify
the authenticity of $\hat{\mathbf{X}}$.  This latter step seems
somewhat odd.

\section{An asymmetric Tardos code construction}
\label{sec:Construction}

In previous asymmetric fingerprinting schemes, it is up to the Buyer
to generate his or her fingerprint.  The Buyer then sends a commitment
to the Provider, which prevents the Buyer from changing the fingerprint during the
protocol. Unfortunately, this cannot be done with a Tardos code since the
fingerprint must follow a given statistical distribution controlled by
$\mathbf{p}$, and $\mathbf{p}$ is only known to the Provider. This section
proposes a solution to this problem, which consists of two phases.  We first review its main building blocks.

\subsection{Building blocks}

There are two key building blocks to the proposed protocol.  The first
is a block involving encryption primitives, while the second involves
double-blind random selection.

\subsubsection{Encryption Primitives}
We need two cryptographic primitives: a regular symmetric cryptosystem
\textbf{E} (e.g. AES) and a commutative encryption scheme \textbf{CE}
(e.g. in~\cite{bao_icisc_00,bao_icics_01}). This latter
primitive has the following property. For every key $k_1$ and $k_2$, and
for every message $m$, ciphering twice with $k_1$ and then $k_2$, or
$k_2$ and then $k_1$ leads to the same result:
\begin{equation}
 \textbf{CE}(k_1, \textbf{CE}(k_2,m)) = \textbf{CE}(k_2, \textbf{CE}(k_1,m)).
\end{equation}

\subsubsection{Pick a card, any card!}
\label{subsec:card}

Here we introduce a double-blind random selection protocol between two entities $\mathtt{A}$ and $\mathtt{B}$, based on~\cite{bao_icisc_00}. 
Let $\{O_k\}_{k=1}^{N}$ be a list of $N$ objects offered by
entity $\mathtt{A}$. We now explain how entity $\mathtt{B}$
selects an item from this list without actually seeing the list and
entity $\mathtt{A}$ does not know which item entity $\mathtt{B}$ picked.

Entity $\mathtt{A}$ chooses $N$ secret keys for the \textbf{E} cryptosystem
called $\{K_k\}_{k=1}^{N}$ and computes the cipher texts
$C_{k}=\textbf{E}(K_k, O_k)$.  {Entity} $\mathtt{A}$ also chooses a secret key
$S$ for the \textbf{CE} cryptosystem and encrypts the previous keys such that
$D_{k}=\textbf{CE}(S,K_{k})$.  He sends $\mathtt{B}$ the lists
$\mathcal{C}=\{C_{k}\}_{k=1}^{N}$ and
$\mathcal{D}=\{D_{k}\}_{k=1}^{N}$.  {Entity} $\mathtt{B}$ chooses an index
$k\in[N]$ (with the notation $[N]=\{1,\ldots,N\}$), a secret key $R$
for the \textbf{CE} cryptosystem, and sends $\mathtt{A}$ the cipher
$U_k=\textbf{CE}(R,D_{k})$.  Entity $\mathtt{A}$ decrypts $U$ with his key $S$
and sends $\mathtt{B}$ the result. Thanks to the commutative property,
this message indeed equals $\textbf{CE}(R,K_{k})$, which $\mathtt{B}$
is able to decrypt thanks to his/her key $R$. The result is the key
$K_{k}$ which deciphers $C_{k}$ onto the object $O_{k}$.

\begin{figure*}
  \centering \input{pickacard_fing_detail4.pstex_t}
  \caption{Generation of a fingerprint bit. }
  \label{fig:test}
\end{figure*}
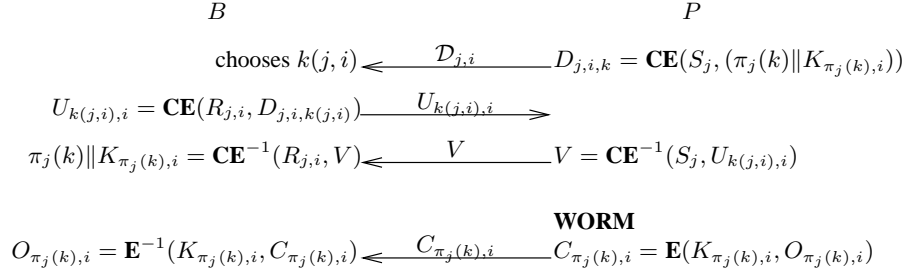

\subsection{Phase 1: Generation of the fingerprint}
\label{subsec:Build}

We use the above protocol $m$ times to generate the fingerprint of the
$j$-th Buyer $\mathbf{X}_j=(X_{j,1},\ldots,X_{j,m})$.  In this
generation phase, $\mathtt{A}$ is the Provider, and $\mathtt{B}$ is
Buyer $j$. The Provider generates a secret vector $\mathbf{p}$ for a
Tardos code. Each $p_{i}$ is quantized such that $p_{i}=L_{i}/N$ with
$L_{i}\in[N-1]$.

For a given index $i$, the objects are the concatenation of a binary
symbol and a text string. There are only two versions of an object in
list $C_i$. For $L_{i}$ objects, $O_{k,i} = (1\|\mathtt{ref}_{1,i})$,
and $O_{k,i}=(0\|\mathtt{ref}_{0,i})$ for the $N-L_i$ remaining ones.
The use of the text strings $\{\mathtt{ref}_{X,i}\}$ depends on the
content distribution mode as detailed in Sec.~\ref{sec:Embed}.  The
object $O_{k,i}$ is encrypted with key $K_{k,i}$ and stored in the
list $\mathcal{C}_i=\{C_{k,i}\}_{k=1}^{N}$. There are thus as many
different lists $\mathcal{C}_i$ as the length $m$ of the
fingerprint. These lists are published in a public Write Once Read
Many (WORM) directory~\cite{Oprea:2009fk} whose access is granted to
all users. As explicitly stated in its name, nobody can modify or
erase what has been put the first time in a WORM directory; beside,
anybody can check its integrity.

On the contrary, the $\mathcal{D}$-lists are made specific to a given
Buyer $j$. The provider picks a secret key $S_{j}$ and a permutation
$\pi_{j}(.)$ over $[N]$.
This Buyer is proposed a list $\mathcal{D}_{j,i}$ of $N$ items as
$D_{j,i,k}=\textbf{CE}(S_j,(\pi_{j}(k)\|K_{\pi_{j}(k),i}))$.
Therefore, the lists $\mathcal{C}_i$ are common for all users, whereas
the lists $\mathcal{D}_{j,i}$ are specific to Buyer $j$. We have
introduced here a slight change wrt to protocol~\ref{subsec:card},
i.e. the permutation $\pi_{j}$ whose role is explained below.  Buyer
$j$ chooses a secret $R_{j,i}$ and one object in the list, say the
$k(j,i)$-th object. He/she sends the corresponding ciphertext
$U_{k(j,i),i}=\textbf{CE}(R_{j,i},D_{j,i,k(j,i)})$ decrypted by the
provider with $S_j$ and sent back to the Buyer who, at the end, gets
the index $\ind(j,i)=\pi_{j}(k(j,i))$ and the key $K_{\ind(j,i),i}$,
which grants him/her the access to the object $O_{\ind(j,i),i}$, store encrypted in the WORM.
It contains the symbol $b_{\ind(j,i),i}$. This will be the value of the
$i$-th bit of his/her fingerprint, $X_{j,i} = b_{\ind(i,j),i}$, which
equals `1' with probability $p_i$.

The provider keeps in a log file the values of $S_j$ and
$U_{k(j,i),i}$, the user keeps $R_{j,i}$ in his/her records.

\subsection{Phase 2: Disclosure of the halfword}
\label{subsec:halfword}

For a more practical accusation process (see Sec.~\ref{sec:Accuse}),
the Provider will order Buyer $j$ to reveal $m_h<m$ bits of his
fingerprint (phase 1 has been completed). This is done in order to build the so-called
halfword~\cite{PfitzmannAsym} allowing the Provider to list a bunch of
suspected users to be forwarded to the judge (See
Sec.~\ref{sec:Accuse}).  The following facts must be enforced: Buyer
$j$ doesn't know which bits of his/her fingerprint are disclosed, and
the Provider asks for the same bit indices to all the users.
  
Again, we propose to use the double-blind random selection protocol of Sec.~\ref{subsec:card}.
Now, Buyer $j$ plays the role of $\mathtt{A}$, and the Provider the role of
$\mathtt{B}$, $N=m$, and object
$O_{i}=(R_{i,j}\|\mathtt{alea}_{i,j})$. These items are the $m$ secret
keys selected by Buyer $j$ during Sec.~\ref{subsec:Build} concatenated
with random strings $\mathtt{alea}_{i,j}$ to be created by Buyer
$j$. This alea finds its use during the personalization of the content
(see Sec.~\ref{sec:Embed}). Following the protocol, the Provider
selects $m_h$ such object. The decryption of message $U_{k(i,j),j}$
received during the construction phase of Sec.~\ref{subsec:Build}
thanks to the disclosure of the key $R_{i,j}$ yields $D_{i,j,k(i,j)}$
which in turn decrypted with key $S_j$ provides the index of the
selected object, otherwise the protocol stops. This prevents a
colluder from denying the symbol of his fingerprint and from copying
the symbol of an accomplice. At the end, the Provider learns which
item was picked by Buyer $j$ at index $i$. Therefore, he/she ends up with
$m_h$ couples $(X_{j,i},\mathtt{alea}_{k(i,j),i})$ associated to a
given Buyer $j$.

Thanks to this second part of our protocol, the Provider discloses $m_h$ bits of the
fingerprints without revealing any knowledge about the others, and Buyer
$j$ doesn't know which bits of his fingerprint were disclosed even if
the Provider always chooses the same indices from a user to another.
Of course, Buyer $j$ refuses to follow this part of the protocol for more than $m_h$ objects.

\section{Other implementation details}
\label{sec:wmaccuse}

At this point, we have both introduced a new attack and a new
asymmetric fingerprinting algorithm that are both specific to
Tardos codes.  The astute reader will be aware the our asymmetric
fingerprinting protocol does not constitute a complete system.  Here
we briefly touch up on other implementation issues. 

\subsection{Watermarking}
\label{sec:Embed}

First, we need an algorithm so that the
Provider sends the Buyer a copy of the Work with his/her fingerprint embedded,
given the Provider does not know this fingerprint.
There exist buyer-seller
protocols for embedding a sequence $\mathbf{X}_{j}$ into a content
$c_{o}$ without disclosing $\mathbf{X}_{j}$ to the seller and $c_{o}$
to the buyer. They are based on homomorphic encryption scheme and work
with some specific implementations of spread
spectrum~\cite{Kuri08Spread} or Quantization Index Modulation
watermarking~\cite{Deng:2009fk}.  The reader is directed to~\cite{Kuri08Spread,Deng:2009fk} for further details.  These methods
can be adapted to embed the Tardos codes, but due to space limitations,
a brief sketch of the adaptation of \cite{Deng:2009fk} is presented hereafter.

 We adapt the secure embedding proposed in the last cited work as
 follows. Let $\mathbf{c}_i^{(0)}=(c^{(0)}_{i,1},\ldots,c^{(0)}_{i,Q})$
 be the $Q$ quantized components (like pixels, DCT coefficients, portion of streams etc) of the $i$-th content block watermarked with symbol `0'
 (resp. $\mathbf{c}_i^{(1)}$ with symbol `1'). Denote $\mathbf{d}_i =
 \mathbf{c}_i^{(1)}-\mathbf{c}_i^{(0)}$. Assume as
 in~\cite[Sect. 5]{Deng:2009fk}, an additive homomorphic and
 probabilistic encryption $E[.]$ such as the Pallier cryptosystem. Buyer
 $j$ has a pair of public/private keys $(pk_{j},sk_j)$ and sends
 $(E_{pk_{j}}[X_{j,1}],\ldots,E_{pk_{j}}[X_{j,m}])$. The provider sends
 him/her the ciphers
 $$
 E_{pk_{j}}[c^{(0)}_{i,\ell}].E_{pk_{j}}[X_{j,i}]^{d_{i,\ell}},\,\forall
 (i,\ell)\in[m]\times[Q].
 $$
 Thanks to the homomorphism, Buyer $j$ decrypts this with $sk_j$ into
 $c^{(0)}_{i,\ell}$ if $X_{j,i}=0$, $c^{(1)}_{i,\ell}$ if $X_{j,i}=1$.
 Since $X_{j,i}$ is constant for the $Q$ components of the $i$-th
 block, a lot of bandwidth and computer power will be saved with a
 composite signal representation as detailed
 in~\cite[Sect. 3.2.2]{Deng:2009fk}.

A crucial step in these buyer-seller protocols is to prove to the
seller that what is sent by the Buyer is indeed the encryption of bits,
and moreover bits of the Buyer's fingerprint.  To do so usually
involves complex zero-knowledge subprotocols~\cite{Kuri08Spread,Deng:2009fk}.
We believe we can avoid
this complexity by taking advantage of the fact that
the Provider already knows some bits of the fingerprint
$\mathbf{X}_{j}$, i.e. those
belonging to the halfword (see
Sect.~\ref{subsec:halfword}), and the Buyers do not know the indices of these bits. Therefore, in $m_v$ random indices of
the halfword, the Provider asks the Buyer $j$ to open his/her
commitment. For one such index $i_{v}$, Buyer $j$ reveals the random
value $r_{i_v}$ of the probabilistic Pallier encryption (with the notation of~\cite{Deng:2009fk}). The Provider
computes $g^{X_{j,i_v}}h^{r_{i_v}}\mod N$ and verifies it equals the
$i_v$-th cipher, which Buyer $j$ pretended to be
$E_{pk_{j}}[X_{j,i}]$. 

One drawback of this simple verification scheme
is that the Buyer discovers $m_v$ indices of the halfword. This may
give rise to more elaborated collusion attacks.  For example, Buyer $j$, as a
colluder, could try to
enforce $Y_{i_v} \neq X_{j,i_v}$ when attempting to forge a pirated
copy.  Further discussion of this is beyond the scope of this paper.

This approach may also introduce a threat to the Buyer.  An untrustworthy Provider can ask to open
the commitments of non-halfword bits in order to disclose bits he/she
is not supposed to know. For this reason, the Provider needs to send
$\mathtt{alea}_{k(i_v,j),i_v}$ as defined in
Sec.~\ref{subsec:halfword} to show Buyer $j$ that his/her verification occurs on a halfword bit.

\subsection{The accusation procedure}
\label{sec:Accuse}

When an unauthorized copy is found, 
the  Provider decodes the watermark and extracts the sequence
$\mathbf{Y}$ from the pirated content. The Provider computes the
halfscores by applying Eq.~\eqref{eq:AccusSum} only on the
halfwords. This produces a list of suspects, e.g. those users whose score
is above a threshold, or those users with the highest scores.

Of course, this list cannot be trusted, since the Provider may be
untrustworthy.   
The list is therefore sent to a third party, referred to as the Judge,  who first verifies the computation of
the halfscores. 
If different values are found, the Provider is
black-listed. Otherwise, the Judge computes the scores of the full
fingerprint.

To do so, the Judge needs the secret $\mathbf{p}$: he/she
asks the Provider for the keys $\{K_{k,i}\}$, $\forall
(k,i)\in[N]\times[m]$ and thereby obtains from the WORM all the objects
$\{O_{k,i}\}$, and therefore the true values of $(p_1,\ldots,p_m)$. 
The Judge must also request suspected Buyer $j$ for the keys $R_{j,i}$
in order to decrypt the messages $U_{k(j,i),i}$ in $D_{i,j,k(i,j)}$ which
reveal which object Buyer $j$ picked during the $i$-th round of
Sec.~\ref{subsec:Build} and whence $X_{j,i}$. 
Finally, the Judge
accuses the user whose score over the full length fingerprint is above a
given threshold (related to a probability of false alarm).

\subsection{Security}
Suppose first that the Provider is honest and denote by $c$ the
collusion size. A reliable tracing capability on the halfwords is
needed to avoid false alarms. Therefore, as proven by G. Tardos, $m_h=O(c^2\log
n\epsilon^{-1})$, where $\epsilon$ is the probability of suspecting
some innocent Buyers. Moreover, successful collusions are avoided if
there are secret values such that $p_i<c^{-1}$ or
$p_i>1-c^{-1}$(see~\cite{Furon2009:Worst}).  Therefore, $N$ should be
sufficiently big, around a hundred, to resist against collusion of size
of the order of ten. 
During the generation of the fingerprint in
Sec.~\ref{subsec:Build}, permutation $\pi_{j}(.)$ makes sure that Buyer
$j$ randomly picks up a bit `1' with probability $p_{i}=L_{i}/N$ as
needed in the Tardos code. In particular, a colluder cannot benefit
from the discoveries made by his accomplices.

We now analyze why colluders would cheat during the watermarking
of their version of the Work described in Sec.~\ref{sec:Embed}.
By comparing their fingerprints, they see indices where they all have the same symbols,
be it `0' or `1'. As explained in the introduction, they won't be able to alter those bits in the tampered fingerprint except if they cheat during the watermarking: 
If their fingerprint bits at index $i$ all equal `1', one of them must pretend he/she has a `0' in this position. If they succeed to do so for all these positions, they will able to forge a pirated copy with a null fingerprint for instance. 

How many times do the colluders need to cheat?
With probability $p_i^c$ (resp. $(1-p_i)^c$), they all have bit `1'
(resp. `0') at index $i$. Thus, there are on average $m_c(c)=m\int_t^{1-t}
(p^c+(1-p)^c)f(p)dp$ such indices. The Provider asks for a bit verification with probability $m_v/m_h$. The probability of a
successful attack for a collusion of size $c$ is therefore
$(1-m_v/m_h)^{m_c(c)}$.
Our numerical simulations have shown that $m_v$ shouldn't be more than 50 bits for typical code length and collusion size below a hundred. Thus, $m_v$ is well below $m_h$. 

Suppose now that the Provider is dishonest. The fact that the $m$
lists $\mathcal{C}_{i},\,\forall i\in[m]$ are public and not
modifiable prevents the Provider from altering them for a specific
Buyer in order to frame him/her afterwards. Moreover, it will raise the
Judge's suspicion if the empirical distribution of the $p_i$ is not
close to the pdf $f$.  Yet, biases can be introduced on the probabilities
for the symbols of the colluders' fingerprint only if there is a
coalition between them and the untrustworthy Provider. For instance, the
Provider can choose a permutation such that by selecting the first
item (resp. the last one) in the list $\mathcal{D}_{j,i}$ an accomplice colluder
is sure to pick up a symbol `1' (resp. `0'). This ruins the tracing
property of the code, but this does not
allow the Provider to frame an innocent.  First, it is guaranteed that
$\mathbf{p}$ used in Eq.~\ref{eq:AccusSum} is the one which generated
the code. Second, the Provider and his accomplices colluders must ignore a significant part of the fingerprints of innocent Buyers. To this end, $m-m_h$ must also be in order of $O(c^2\log n\epsilon^{-1})$. If this holds, the Judge is able to take a reliable decision while discarding the halfword part of the fingerprint. Consequently, $m\approx 2m_h$, our protocol has doubled the typical code length, which is still in $O(c^2\log n\epsilon^{-1})$.

\section{Discussion and summary}
\label{sec:discussions}

Tardos codes are currently the state-of-the-art in collusion-resistant
fingerprinting. However, the previous asymmetric fingerprint protocols 
cannot be applied to this particular construction. There are mainly two difficulties.
First, the Buyer has to generate his/her secret fingerprint but according to vector $\mathbf{p}$, which is kept secret by the Provider. Second, the vector $\mathbf{p}$ used in the accusation process must be the same as the one which generated the fingerprints.

We have proposed a new asymmetric fingerprinting protocol dedicated to Tardos codes.  We believe that this is the first such protocol, and that it is practically efficient.

The construction of the fingerprints and their embedding within pieces of Work do not
need a trusted third party. 

Note, however, that during the accusation stage, a
trusted third party is necessary like in any asymmetric fingerprinting scheme we are aware of. Further work is needed to determine
if such a third-party can be eliminated.  In particular, we anticipate
that some form of secure multi-party computation can be applied.

Other extensions to this work include (i) non-binary Tardos codes, and
(ii) implementation on compliant consumer devices such as Blu-Ray
players.  We also plan to develop this as part of future work.

\bibliographystyle{IEEEbib}
\bibliography{mabiblio}

\end{document}

%% file: pickacard_fing_detail4.pstex_t
\begin{picture}(0,0)%
\includegraphics{pickacard_fing_detail4.pstex}%
\end{picture}%
\setlength{\unitlength}{3947sp}%
\begingroup\makeatletter\ifx\SetFigFont\undefined%
\gdef\SetFigFont#1#2#3#4#5{%
  \reset@font\fontsize{#1}{#2pt}%
  \fontfamily{#3}\fontseries{#4}\fontshape{#5}%
  \selectfont}%
\fi\endgroup%
\begin{picture}(3030,1737)(3586,-3139)
\put(4489,-1861){\makebox(0,0)[rb]{\smash{{\SetFigFont{12}{14.4}{\rmdefault}{\mddefault}{\updefault}{\color[rgb]{0,0,0}\begin{small}chooses $k(j,i)$\end{small}}%
}}}}
\put(4513,-2161){\makebox(0,0)[rb]{\smash{{\SetFigFont{12}{14.4}{\rmdefault}{\mddefault}{\updefault}{\color[rgb]{0,0,0}\begin{small}$\U$\end{small}}%
}}}}
\put(5101,-1825){\makebox(0,0)[b]{\smash{{\SetFigFont{12}{14.4}{\rmdefault}{\mddefault}{\updefault}{\color[rgb]{0,0,0}\begin{small}$\listedrow$\end{small}}%
}}}}
\put(5101,-2125){\makebox(0,0)[b]{\smash{{\SetFigFont{12}{14.4}{\rmdefault}{\mddefault}{\updefault}{\color[rgb]{0,0,0}\begin{small}$\Urow$\end{small}}%
}}}}
\put(5713,-2461){\makebox(0,0)[lb]{\smash{{\SetFigFont{12}{14.4}{\rmdefault}{\mddefault}{\updefault}{\color[rgb]{0,0,0}\begin{small}$V=\textbf{CE}^{-1}(S_j,\Urow)$\end{small}}%
}}}}
\put(5101,-2425){\makebox(0,0)[b]{\smash{{\SetFigFont{12}{14.4}{\rmdefault}{\mddefault}{\updefault}{\color[rgb]{0,0,0}\begin{small}$V$\end{small}}%
}}}}
\put(4513,-2461){\makebox(0,0)[rb]{\smash{{\SetFigFont{12}{14.4}{\rmdefault}{\mddefault}{\updefault}{\color[rgb]{0,0,0}\begin{small}$\decrypt$\end{small}}%
}}}}
\put(5101,-3025){\makebox(0,0)[b]{\smash{{\SetFigFont{12}{14.4}{\rmdefault}{\mddefault}{\updefault}{\color[rgb]{0,0,0}\begin{small}$\listcrow$\end{small}}%
}}}}
\put(5713,-1861){\makebox(0,0)[lb]{\smash{{\SetFigFont{12}{14.4}{\rmdefault}{\mddefault}{\updefault}{\color[rgb]{0,0,0}\begin{small}$\listed$\end{small}}%
}}}}
\put(5713,-3061){\makebox(0,0)[lb]{\smash{{\SetFigFont{12}{14.4}{\rmdefault}{\mddefault}{\updefault}{\color[rgb]{0,0,0}\begin{small}$\listc$\end{small}}%
}}}}
\put(4489,-3061){\makebox(0,0)[rb]{\smash{{\SetFigFont{12}{14.4}{\rmdefault}{\mddefault}{\updefault}{\color[rgb]{0,0,0}\begin{small}$\decryptC$\end{small}}%
}}}}
\put(5713,-2881){\makebox(0,0)[lb]{\smash{{\SetFigFont{12}{14.4}{\rmdefault}{\mddefault}{\updefault}{\color[rgb]{0,0,0}\begin{small}\textbf{WORM} \end{small}}%
}}}}
\put(3601,-1561){\makebox(0,0)[b]{\smash{{\SetFigFont{12}{14.4}{\rmdefault}{\mddefault}{\updefault}{\color[rgb]{0,0,0}\begin{small}$\Buyer$\end{small}}%
}}}}
\put(6601,-1561){\makebox(0,0)[b]{\smash{{\SetFigFont{12}{14.4}{\rmdefault}{\mddefault}{\updefault}{\color[rgb]{0,0,0}\begin{small}$\Provider$ \end{small}}%
}}}}
\end{picture}%